\documentstyle[aps,prb]{revtex} 
\begin{document}

\twocolumn[\hsize\textwidth\columnwidth\hsize\csname
@twocolumnfalse\endcsname

\title{Optical absorption and energy-loss spectra of aligned carbon nanotubes}
\author{F. J. Garc\'\i a-Vidal$^1$ and J. M. Pitarke$^2$}
\address{$^1$ Departamento de F\'\i sica Te\'orica de la Materia
Condensada, Facultad
de Ciencias, Universidad Aut\'onoma de Madrid,\\
28049 Madrid, Spain\\
$^2$ Materia Kondentsatuaren Fisika Saila, Zientzi Fakultatea, 
Euskal Herriko Unibertsitatea,\\ 644 Posta kutxatila, 48080 Bilbo, Basque
Country, Spain\\
and\\
Donostia International Physics Center (DIPC) and Centro Mixto
CSIC-UPV/EHU,\\ Donostia, Basque Country, Spain}

\date\today

\maketitle

\begin{abstract}
Optical-absorption cross-sections and energy-loss spectra of aligned multishell
carbon nanotubes are investigated, on the basis of photonic band-structure
calculations. A local graphite-like dielectric tensor is assigned to every
point of the tubules, and the effective transverse dielectric function of the
composite is computed by solving Maxwell's equations in media with tensor-like
dielectric functions. A Maxwell-Garnett-like approach appropriate to the case
of infinitely long anisotropic tubules is also developed. Our full
calculations indicate that the experimentally measured macroscopic dielectric
function of carbon nanotube materials is the result of a strong
electromagnetic coupling between the tubes. An analysis of the electric-field
pattern associated with this coupling is presented, showing that in the
close-packed regime the incident radiation excites a very localized
tangential surface plasmon.
\end{abstract}
]

\narrowtext

\section{Introduction}

The description of the electronic response of carbon
nanotubes\cite{Ijima1,Ebbensen1} has been a challenge for theoretical and
experimental investigations. Various theoretical studies of plasmon excitations
in single-shell carbon nanotubes were
reported,\cite{Lin,Sato,Longe} electron-energy-loss spectra from individual
multishell nanotubes were investigated\cite{Kuzuo,Ajayan3,Bursill,Bonard} by
changing the number of shells, and curvature was found to induce little effect
on the covalent bonding of multishell nanotubes.\cite{Ajayan4} With the
availability of aligned carbon nanotube films,\cite{Ajayan,Heer1} optical
measurements were carried out with polarized light,\cite{Heer1,Heer3}
thereby evaluating the frequency-dependent effective dielectric function of
the composite and showing that carbon nanotubes have an intrinsic and
anisotropic metallic behaviour. Accurate calculations of the effective
dielectric function of densely packed carbon nanotubes, as obtained by solving
Maxwell's equations with the use of tensor-like dielectric functions, have been
carried out only very recently.\cite{fj1,fjnew}

In this paper we report extensive calculations of 
the frequency-dependent effective dielectric function of a composite made
up of aligned carbon nanotubes embedded in an otherwise homogeneous medium. In
section II our effective medium theory is described. We take an electromagnetic
wave normally incident on the structure, and focus on the case of electromagnetic 
waves polarized normal to the cylinders ($p$ polarization).  For this
polarization we also introduce a generalized Maxwell-Garnett (MG)\cite{MG}
effective dielectric function appropriate for anisotropic tubules. Absorption of
$p$ polarized light is found to be sensitive to both the inner cavity of hollow
tubules and the anisotropy, which we first investigate in section IIIA within
the MG approach. In section IIB we focus on the close-packed regime, showing
that the experimentally measured macroscopic dielectric function of aligned
carbon nanotubes is the result of a strong electromagnetic coupling between the
tubes. Calculations of the electric field, the induced charge, and the so-called
energy-loss function, i.e., the imaginary part of the effective inverse
dielectric function, are also presented, for various values of the filling
fraction and the ratio of the internal and external radii of the cylinders. The
main conclusions of our work are addressed in section IV.

\section{Theory}

Take a periodic array of infinitely long multishell nanotubes of inner and
outer radii $r$ and $R$, respectively, arranged in a square array with lattice
constant $a=2\,x\,R$, as shown in Fig. 1. These tubules are assumed to be
embedded in an insulating medium, with a real and positive dielectric constant
$\varepsilon_0$. In the energy range of interest in the interpretation of
absorption cross sections and energy-loss spectra the diameter of typical
multishell carbon nanotubes ($2\,R\sim 10\,{\rm nm}$) is small in comparison to
the wavelength of light, and we also assume that this diameter is large enough
that a macroscopic dielectric function is ascribable to the tubules. For
simplicity, the magnetic permeabilities will be assumed to be equal to unity in
all media.

Planar graphite is a highly anisotropic material, and the dielectric
function is a tensor. This tensor may be diagonalized, by choosing cartesian
coordinates with two of the axes lying in the basal plane and the third axis
being the so-called $c$-axis. One defines the dielectric function
$\varepsilon_\perp(\omega)$ perpendicular to the $c$-axis and the dielectric
function $\varepsilon_\parallel(\omega)$ for the electric field parallel to the
$c$-axis. For carbon nanotubes, we assume full transferability of the dielectric
tensor of planar graphite to the curved geometry of carbon tubules, as suggested
by Lucas {\it et al\,}\cite{Lucas} for the case of multishell fullerenes. Hence,
we simply assign a local graphite-like dielectric tensor to every point inside
the nanotube and outside the inner core, and write
\begin{equation}\label{eq1}
\hat{\varepsilon} (\omega)=\varepsilon_{\perp}(\omega)
({\bf \theta}{\bf \theta}+{\bf z}{\bf z})+
\varepsilon_{\parallel}(\omega){\bf r}{\bf r},
\end{equation}
where ${\bf \theta}{\bf \theta}$, ${\bf z}{\bf z}$, and ${\bf r}{\bf r}$
are the unitary basis vectors of cylindrical coordinates.

In the long-wavelength limit, a composite material may be treated as if it
were homogeneous, with the use of an effective dielectric function
$\varepsilon_{eff}$. The optical absorption cross section of the composite is
then directly given by ${\rm Im}\,\varepsilon_{eff}(\omega)$. Also, for small
values of the dimensionless parameter $qR$ ($qR<1$), $q$ being the momentum
transfer, the energy-loss spectra of a broad beam of swift electrons penetrating
the composite is found\cite{Pitarke1,Pitarke2} to be well described by the $q\to
0$ limit of the imaginary part of the effective dielectric function, i.e., the
so-called energy-loss function, ${\rm Im}[-\varepsilon_{eff}^{-1}(\omega)]$.  

We consider an electromagnetic wave normally incident on the structure, so that
$k_y=k_z=0$. For this propagation direction there are two different values of
$\varepsilon_{eff}(\omega)$ corresponding to $s$ and $p$ polarizations. In the
case of $s$ polarization the electric field is parallel to the cylinders at
every point, and is not modified by the presence of the interfaces. Hence, the
$s$ effective dielectric function of the composite is simply the weighted average
of the dielectric functions of the constituents.\cite{Bergman}

For electromagnetic waves polarized normal to the cylinders ($p$ polarization), the
electric field may be strongly modified by the presence of the interfaces.
An elementary analysis shows that
\begin{equation}\label{eq5}
(\varepsilon_{eff}-\varepsilon_0){\bf
E}=f(\hat\varepsilon-\hat\varepsilon_0)\cdot{\bf
E}_{in},
\end{equation}
where ${\bf E}$ is the average electric field over the composite,
\begin{equation}\label{eq6}
{\bf E}=f{\bf E}_{in}+(1-f){\bf E}_{out},
\end{equation}
${\bf E}_{in}$ and ${\bf E}_{out}$ representing the average electric field
inside
and outside the tubules, respectively, both lying in the plane of
periodicity.

In the case of a single two-dimensional circular inclusion (plain or hollow
cylinder)
embedded in an otherwise homogeneous medium ($f\to 0$), one easily finds
\begin{equation}\label{eq7}
(\hat\varepsilon-\hat\varepsilon_0)\cdot{\bf
E}_{in}=\varepsilon_0\alpha{\bf E},
\end{equation}
and Eq. (\ref{eq5}) then yields
\begin{equation}\label{eq8}
\varepsilon_{eff}=\varepsilon_0\,\left(1+f\,\alpha\right),
\end{equation}
where $\alpha$ represents the in-plane dipole polarizability per unit
volume.

As long as the composite is made of a sparse ($f<<1$) distribution of
cylinders the presence of multipolar modes can be neglected,\cite{Pitarke3}
and the
interaction between the cylinders can be introduced by simply replacing the
average
electric field
${\bf E}$ in Eq. (\ref{eq7}) by ${\bf E}_{out}$. Then, with the aid of Eqs.
(\ref{eq5}) and (\ref{eq6}), one finds
\begin{equation}\label{eq11}
\varepsilon_{eff}=\varepsilon_0\,\left(1+f\,{\alpha\over
1-f\,L\,\alpha}\right)
\end{equation}
and
\begin{equation}\label{eq12}
\varepsilon_{eff}^{-1}=\varepsilon_0^{-1}\,\left(1-f\,{\alpha\over
1+f\,L\,\alpha}\right),
\end{equation}
with the geometrical factor $L=1/2$. For cylindrical geometry,\cite{Lucas2}
\begin{eqnarray}\label{eq9}
&&\alpha={2\over(1-\rho^2)}\cr\cr
&&\times{(\varepsilon_\parallel\Delta-\varepsilon_0)
(\varepsilon_\parallel\Delta+\varepsilon_0)-(\varepsilon_\parallel\Delta-\varepsilon_0)
(\varepsilon_\parallel\Delta+\varepsilon_0)
\rho^{2\Delta}\over
(\varepsilon_\parallel\Delta+\varepsilon_0)(\varepsilon_\parallel\Delta+\varepsilon_0)
-(\varepsilon_\parallel\Delta-\varepsilon_0)(\varepsilon_\parallel\Delta-\varepsilon_0)
\rho^{2\Delta}},
\end{eqnarray}
where
\begin{equation}\label{eq10}
\Delta=\sqrt{\varepsilon_\perp/\varepsilon_\parallel}.
\end{equation}
Eq. (\ref{eq9}) with $\rho=0$ and
$\varepsilon_\parallel(\omega)=\varepsilon_\perp(\omega)$ reduces
to the well-known polarizability per unit volume of a plain isotropic
cylinder. 

Eq. (\ref{eq11}) is a generalization, appropriate for anisotropic tubules,
of the well-known MG effective dielectric function derived by
Maxwell-Garnett for a system of spherical particles.\cite{MG}

In the non-sparse or packed regime, where the presence of multipolar modes
cannot be neglected, the inclusion of the full electromagnetic interaction
between the tubules is unavoidable. In order to compute, with full inclusion of
this interaction, the effective dielectric function of our periodic system, we
have followed the method developed in Refs.\onlinecite{Pendry1}
and\onlinecite{Pendry2} for the calculation of dispersion relationships
$k(\omega)$ of Bloch waves in structured materials with tensor-like dielectric
functions.

In the long-wavelength limit the composite material supports, for each
polarization, only two degenerate electromagnetic Bloch waves with vectors $k$ and
$-k$ for which $k(\omega)$ roughly follows the dispersion relation of free light.
Hence, in this limit the composite may be treated as if it were homogeneous, with
an effective transverse dielectric function
\begin{equation}\label{eq3}
\varepsilon_{eff}(\omega)={k^2(\omega)c^2\over\omega^2},
\end{equation}
where $c$ represents the speed of light.

\section{Results and discussion}

We consider a periodic array of hollow multishell carbon nanotubes (see
Fig. 1) with
$\varepsilon_0=1$, and take the principal dielectric functions
$\varepsilon_\perp(\omega)$ and
$\varepsilon_\parallel(\omega)$ of graphite from Ref.\onlinecite{Palik}.
The real and
imaginary parts of these components, as well as their corresponding
energy-loss
functions, are represented in Fig. 2 for energies up to $8\,{\rm eV}$. In
this
energy range, optical transitions mainly involve the $\pi$ bands arising
from the
atomic $2p_z$ orbitals. The peak in
${\rm Im}\,\varepsilon_\perp(\omega)$ at $\sim 4.6\,{\rm eV}$ is associated with
the maximum
in the joint density of states of the $\pi$ valence and conduction bands.
The
so-called $\pi$ plasmon at $\sim 7\,{\rm eV}$, where ${\rm
Re}\,\varepsilon_\perp(\omega)$
and ${\rm Im}\,\varepsilon_\perp(\omega)$ are small, is due to a $\pi-\pi^*$ 
interband transition.\cite{Taft}

For energies over $\sim 5\,{\rm eV}$, the principal dielectric function
$\varepsilon_\perp(\omega)$ is close to that of a free-electron gas. Thus,
for these
energies $\pi$ electrons within each shell act as if they were free.
However, for free-electron-like materials ${\rm
Re}\,\varepsilon_\perp(\omega)$
approaches large negative values with decreasing energy, whereas for
graphite the
plasmon region with ${\rm Re}\,\varepsilon_\perp(\omega)$ negative is
bounded from
below and from above at about $5$ and $7\,{\rm eV}$.

Calculations of the $s$ component of the effective dielectric function of an
array of coaxial nanotubes were reported in Ref.\onlinecite{fjnew} for
various values of the filling fraction $f$ of the tubes, showing that they
roughly reproduce the experimentally determined ${\rm Im}\epsilon_{eff}(\omega)$
for $f\sim 0.5$.

Here we focus on the case of electromagnetic waves polarized normal to the cylinders ($p$
polarization), which we first investigate within the MG approach.

\subsection{Maxwell-Garnett approach}

First of all, we ignore the anisotropy and simply assume that
$\varepsilon(\omega)=\varepsilon_\parallel(\omega)=\varepsilon_\perp(\omega)$. In
this case, the in-plane
dipole polarizability of Eq. (\ref{eq9}) can be expressed in the form of a
spectral
representation:
\begin{equation}\label{eq1p0}
\alpha=-\left({B_+\over u-m_-}+{B_-\over u-m_+}\right),
\end{equation}
where $u$ is the spectral variable
\begin{equation}\label{eq3p}
u=\left(1-\varepsilon/\varepsilon_0\right)^{-1},
\end{equation}
$m_\pm$ are depolarization factors,
\begin{equation}\label{eq4p}
m_\pm={1\over 2}\left(1\pm\rho\right),
\end{equation}
and $B\pm$ are the strengths of the corresponding normal modes,
\begin{equation}\label{eq6p}
B_\pm={1\over 2}.
\end{equation}
The depolarization factors of Eq. (\ref{eq4p}) lie on the
segment [0,1] for all values of $\rho$. Hence, normal modes only occur in
the so-called
plasmon region where the dielectric function $\varepsilon(\omega)$ is
negative or equal
to zero. For graphite, this plasmon region is bounded from below and from
above at
about $5$ and $7\,{\rm eV}$ where the spectral variable $u$ is never
smaller than
$u\sim 0.2$. As the ratio between the inner and outer radii of the tubes
goes to unity
($\rho\to 1$), one easily finds
\begin{equation}\label{eq7p}
\alpha={1\over 2}\left(\varepsilon-\varepsilon^{-1}\right).
\end{equation}

In the case of an isolated plain ($\rho=0$) or hollow ($\rho\neq 0$)
cylinder, both
${\rm Im}\,\varepsilon_{eff}(\omega)$ and ${\rm
Im}[-\varepsilon_{eff}^{-1}(\omega)]$ are proportional
to the imaginary part of the in-plane dipole polarizability $\alpha$, and
satisfy the
relation
\begin{equation}
{\rm Im}\left[-\varepsilon_{eff}^{-1}\right]=\epsilon_0^{-2}\,{\rm
Im}\,\varepsilon_{eff}.
\end{equation}
Hence, the absorption and energy-loss spectra of isolated plain ($\rho=0$)
cylinders in vacuum ($\varepsilon_0=1$) exhibit a strong maximum at $\sim
6.5\,{\rm
eV}$ where
$u=1/2$ and
$\varepsilon(\omega)=-1$, as shown in Fig. 3. For hollow tubes ($\rho\neq
0$), there
are two distinct dipolar modes with either tangential or radial symmetry,
at
$u=m_-<1/2$ and $u=m_+>1/2$, respectively, similar to those present in the
case of a thin planar film\cite{Ritchie} and a spherical
shell.\cite{Lucas0} The
corresponding peak heights in ${\rm Im}\,\alpha(\omega)$ are easily found to be
$H\,B_\pm/\sqrt{m_\pm}$, $H$ representing the peak height in the bulk
energy-loss
function, ${\rm Im}[-\varepsilon^{-1}(\omega)]$. Thus, the tangential mode at
$u=(1-\rho)/2$ appears to be more pronounced than the radial mode at
$u=(1+\rho)/2$
(see Fig. 3). For $\rho=0.2$ these modes are not resolved, due to the
smoothing effect
of the large damping associated with non-negligible values of
${\rm Im}\,\varepsilon_\perp(\omega)$, and the effect of the empty core is simply
to redshift
and soften this combined plasmon mode. For values of $\rho$ in the range
$0.2<\rho<0.8$ the tangential and radial plasmons are clearly identified.
However, for
larger values of $\rho$ the tangential resonance condition ($u\to 0$) is
never
satisfied. Furthermore, in the limit as $\rho\to 1$ the effective
dielectric function
is given by Eq. (\ref{eq7p}), thus showing the radial plasmon resonance at
$\sim
7\,{\rm eV}$, where $u=1$ and $\varepsilon(\omega)=0$, and the
characteristic peak at
$\sim 4.6\,{\rm eV}$ associated with the maximum in ${\rm
Im}\,\varepsilon(\omega)$.

Still ignoring the anisotropy, the MG effective dielectric function and
inverse
dielectric function of Eqs. (\ref{eq11}) and (\ref{eq12}) can also be
expressed in
the form of a spectral representation:
\begin{equation}\label{eq1p}
\varepsilon_{eff}=\varepsilon_0\left[1-f\left({B_+\over u-m_-}+{B_-\over
u-m_+}\right)\right]
\end{equation}
and
\begin{equation}\label{eq2p}
\varepsilon_{eff}^{-1}=\varepsilon_0\left[1+f\left({B_-\over
u-n_-}+{B_+\over
u-n_+}\right)\right],
\end{equation}
where
\begin{equation}\label{eq4pp}
m_\pm={1\over 2}\left(1-{1\over 2}f\pm{1\over 2}\sqrt{f^2+4\rho^2}\right),
\end{equation}
\begin{equation}\label{eq5pp}
n_\pm={1\over 2}\left(1+{1\over 2}f\pm{1\over 2}\sqrt{f^2+4\rho^2}\right),
\end{equation}
and
\begin{equation}\label{eq6pp}
B_\pm={1\over 2}\,{\sqrt{f^2+4\rho^2}\pm f\over \sqrt{f^2+4\rho^2}},
\end{equation}
the mode strengths ($B_-\le 1/2$ and $B_+\ge 1/2$) adding up to unity, i.e.,
$B_++B_-=1$. Also,
$m_-,n_-\le 1/2$ and $m_+,n_+\ge 1/2$.

In Fig. 4, we show the surface-mode positions $m_\pm$ and $n_\pm$ of Eqs.
(\ref{eq4pp}) and (\ref{eq5pp}), as a function of $\rho$, for various
values of the
ratio
$x$ between the lattice constant and the outer diameter of the cylinders. In the
dilute limit
($x\to\infty$), both $m_\pm$ and $n_\pm$ coincide with the depolarization
factors of
Eq. (\ref{eq4p}) entering the spectral representation of the
polarizability. As $x$
decreases, for each value of $\rho$ these surface modes split into four
distinct
modes, $m_\pm$ and $n_\pm$, which satisfy the simple relation:
\begin{equation}
n_\pm=1-m_\mp.
\end{equation}

The strengths $B_\pm$, as obtained from Eq. (\ref{eq6pp}), are plotted in
Fig. 5, as a
function of $\rho$, also for various values of the ratio $x$ between the
lattice
constant and the outer diameter of the cylinders. As in the case of the
depolarization
factors, in the dilute limit ($x\to\infty$) the strengths $B_\pm$ coincide
with the
strengths
$B_\pm=1/2$ entering the spectral representation of the polarizability. As
$x$
decreases, these strengths are different, especially for small values of
$\rho$,
showing that the strongest modes occur at $u=m_-$ and $u=n_+$ with the
depolarization
factors $m_-$ and $n_+$ lying on the segments [0,(1-f)/2] (low-energy
resonance) and
[(1+f)/2,1] (high-energy resonance), respectively (see Fig. 4). The modes
at
$u=m_-$ and
$u=n_+$, which have no strength for $\rho=0$, lie on the segments [1/2,1]
and [0,1/2],
respectively. For $x\to 1$, one easily finds
$B_+=1/(1+\rho^2)$ and
$B_-=\rho^2/(1+\rho^2)$, thereby showing that modes occurring at $u=m_-$
and $u=n_+$ dominate. Hence, when cylinders are touching and for most
values of
$\rho$ there is a single dominating resonance condition at $u\to 0$ (low
energy) and
$u\to 1$ (high energy) in ${\rm Im}\,\varepsilon_{eff}(\omega)$ and ${\rm
Im}[-\varepsilon_{eff}^{-1}(\omega)]$, respectively, the strength of the
remaining resonance
being negligible.

From Eqs. (\ref{eq1p}) and (\ref{eq2p}), we have calculated ${\rm Im}\,\varepsilon_{eff}(\omega)/f$ and ${\rm 
Im}[-\varepsilon_{eff}^{-1}(\omega)]/f$ for various values of $x$ and the ratio $\rho$ between the inner and outer radii of the cylinders. While for $x=3$ both
${\rm Im}\,\varepsilon_{eff}(\omega)/f$ and
${\rm Im}[-\varepsilon_{eff}^{-1}(\omega)]/f$
are found to nearly coincide
with the imaginary part of the polarizability of Eq. (\ref{eq1p0}) (see
Fig. 3), we have found that the trend with decreasing the distance between the cylinders is for the low-energy and high-energy resonances to slightly move from the single-cylinder dipole resonances at
$u=(1\pm\rho)/2$ to lower and
higher
energies, respectively; this is obvious in Fig. 6, where ${\rm Im}\,\varepsilon_{eff}(\omega)/f$ and ${\rm
Im}[-\varepsilon_{eff}^{-1}(\omega)]/f$, as obtained from Eqs. (\ref{eq1p}) and (\ref{eq2p}), have been represented for $x=1.5$ and various values of $\rho$. We also note that as $x$ decreases the low-energy
($u=m_-$) and high-energy ($u=n_+$) modes, whose energy position depends
only weakly
on $\rho$, dominate the optical absorption and energy-loss, respectively.

The role of anisotropy is displayed in Figs 7 and 8. In these
figures, we have
plotted ${\rm Im}\,\alpha(\omega)$ (Fig. 7), ${\rm
Im}\,\varepsilon_{eff}(\omega)/f$ (Fig. 8), and
${\rm Im}[-\varepsilon_{eff}^{-1}(\omega)]/f$ (Fig. 8), as obtained from
Eqs. (\ref{eq9}),
(\ref{eq11}), and (\ref{eq12}) with full inclusion of the anisotropic
dielectric function of graphite and for various values of
$\rho$, as in Figs. 3 and 6. With the presence of anisotropy, the
two-mode
structure exhibited by Eqs. (\ref{eq1p0}), (\ref{eq1p}), and (\ref{eq2p})
is replaced
by a more complicated spectral representation. One sees from Fig. 7 that the
tangential plasmon
peak-position of isolated cylinders remains fairly insensitive to the
anisotropy of
carbon nanotubes. However, its shape drastically changes, as a consequence
of the
presence of a nearly constant and positive dielectric function
$\varepsilon_\parallel$, and the radial resonance condition is not visible.
For $x=3$ the MG dielectric function is found to nearly coincide with the
isolated-cylinder
result, as in the isotropic case. As the distance between the cylinders
decreases, the
impact of the anisotropy is still to soften the resonances. Furthermore,
one sees from
Fig. 8 that now at small values of $\rho$ the low-energy resonance is only
visible in the optical absorption, while the energy-loss only exhibits the
high-energy peak.

\subsubsection{Packed regime}

In this section we present results of our calculation of the effective
dielectric
function of a periodic array of carbon nanotubes, as obtained from Eq.
(\ref{eq3}) with
full inclusion of both the anisotropy and the electromagnetic interaction between
the tubules. All calculations presented here have been found to be insensitive
to the precise value of
the number of mesh points in the unit cell. For metallic structures,
sampling meshes
as large as $180\times 180$ have been found to be required to
provide
well-converged results;\cite{Pitarke2} however, for carbon nanotubes
sampling meshes
of $60\times 60$ have been found to provide well-converged
results, which is 
due to the smoothing effect of the large damping originated with the
presence of
interband transitions in graphite.

In Figs. 9 and 10 we show our full calculations of the $p$ component of the
effective dielectric function of an array of plain
($\rho=0$) carbon nanotubes, as obtained for various
values of the ratio $x$ between the lattice constant and the outer diameter of
the
cylinders: $x=2.0$, $x=1.5$ and $x=1.3$ in Fig. 9, and $x=1.1$ and
$x=1.03$ in Fig. 10. Also plotted in these figures by dotted lines are
calculations of the MG dielectric
function of Eqs. (\ref{eq11}) and (\ref{eq12}). These calculations show
that the
actual effective dielectric function is well described by the MG
approximation in the
low-filling-ratio regime, our full calculations beginning to deviate from
those
obtained within the MG approximation at $x\sim 1.3$. For smaller
concentrations of
graphite, multipolar modes cannot be neglected and the dipole-resonance
positions
necessarily deviate, as discussed in Ref.\onlinecite{Pitarke2}.

One sees from Figs. 9 and 10 that, as within the MG approach, the trend
with
increasing the concentration of tubules is for the actual dipolar peak in
${\rm Im}\,\varepsilon_{eff}(\omega)$ and ${\rm
Im}[-\varepsilon_{eff}^{-1}(\omega)]$ to be
shifted from the isolated-cylinder dipole mode at $\sim 6.5\,{\rm eV}$ to
lower and higher
energies, respectively. When the nanoparticles are brought into
close contact, electromagnetic coupling between them converts the dipolar
surface mode
into a very localized one, trapped in the region between the
nanostructures, and the MG
approximation fails to describe the details of the effective dielectric
function. We
note that multipolar resonances, which are present in the close-contact
regime, are
not visible in the spectra due to the smoothing effect of the large damping
characteristic of graphite. At higher concentrations of tubules, when
graphite forms a connected medium ($x\le
1$), dipolar modes cannot be excited. Hence, optical absorption exhibits a
single
peak originated in the maximum of ${\rm Im}\,\varepsilon(\omega)$ at
$\sim 4.6\,{\rm eV}$,
and the energy-loss function shows a single peak at the bulk plasmon
resonance at $\sim
7\,{\rm eV}$. Besides these peaks, there is a background of unresolved
multipole
contributions to ${\rm
Im}\,\varepsilon_{eff}(\omega)$ and ${\rm
Im}[-\varepsilon_{eff}^{-1}(\omega)]$ in the
plasmon region with energies between $\sim 5$ and $\sim 7\,{\rm eV}$.

Calculations of the $p$ component ($p$ polarization) of the effective dielectric function of a periodic array of
hollow ($\rho\neq 0$) carbon nanotubes were reported in Ref.\onlinecite{fjnew}
for various ratios $\rho$ between inner and outer radii of the tubules. As in
the case of plain cylinders ($\rho=0$), these calculations nearly coincide for
$x=2.0$ with the results obtained within the MG approximation. However, in the
close-packed regime ($x=1.03$) the strong electromagnetic coupling between the
tubes results in non-negligible contributions from multipolar resonances. This
multipolar coupling provokes a redshift and a blueshift of the MG dipolar
resonances that are visible in the optical spectra and the energy loss, i.e.,
the low-energy ($u<1/2$) dipolar mode with tangential symmetry and the
high-energy ($u>1/2$) dipolar mode with radial symmetry. A comparison between
these calculations and the experimentally determined macroscopic $p$ dielectric
function of close-packed carbon nanotubes\cite{Heer1} was also presented in
Ref.\onlinecite{fjnew}, showing an excellent agreement for $\rho=0.6$, which
yields in the close-packed regime ($x=1.03$) a filling fraction
$f\sim 0.5$.

Finally, we look at the {\bf E}-field pattern associated 
with the electromagnetic resonances that are present in the optical absorption
and the energy loss. In Fig. 11, we show detailed pictures of the intensity  of
the electric field and the corresponding charge density generated  by normally
incident $p$-polarized light impinging on our periodic array of hollow 
($\rho=0.6$) carbon nanotubes with
$x=2$. The frequency of the incident radiation has been chosen to be
$\omega_1=5.2\,{\rm eV}$ and $\omega_2=5.9\,{\rm eV}$, corresponding to the
location of resonances in the optical absorption and the energy loss,
respectively. In this figure one clearly sees that the incident radiation is
exciting prototypical dipolar tangential and radial dipolar plasmons at the
surfaces of a nearly isolated hollow cylinder. At $\omega_1=5.2\,{\rm eV}$, the
intensity of the electric field is maximum outside the nanotube and the
induced charge is clearly located on the outer surface of the tubule, showing
a tangential field pattern. At $\omega_2=5.9\,{\rm eV}$, the {\bf E}-field
intensity is maximum in the hollow core of the nanotube and the induced charge
is located on both surfaces of each tubule, though the largest part of it is
located on the inner surface of the tube. These represent the main
characteristics of a radial dipolar mode.

In the case of a close-packed structure, the incident radiation excites a very
localized tangential surface plasmon. This is illustrated in Fig. 12, where we
show pictures of the {\bf E}-field intensity and the corresponding charge
density generated by normally incident $p$-polarized light impinging on our
periodic array of hollow ($\rho=0.6$) carbon nanotubes with $x=1.03$. At
$\omega_1=4.75\,{\rm eV}$, where the optical absorption exhibits a maximum,
the {\bf E}-field intensity is strongly enhanced in the region between the
tubes and the induced charge is clearly localized, showing a strongly
localized tangential field pattern. At $\omega_2=6.1\,{\rm eV}$, where the
energy loss is maximum, one clearly sees that the incident radiation is
exciting a radial surface plasmon which is very similar to that observed in
the case of nearly isolated nanotubes (see Fig. 11).

\section{Summary and conclusions}

We have reported extensive calculations of the effective electronic
response of aligned multishell carbon nanotubes, of interest in the
interpretation of absorption spectra and electron energy-loss experiments. A
local graphite-like dielectric tensor has been assigned to every point of the
multishell tubules, and the effective dielectric function of the composite has
been computed by solving Maxwell's equations in media with tensor-like
dielectric functions. A MG-like approach appropriate to the case of infinitely
long anisotropic tubules has also been developed, showing that MG results are
accurate as long as the distance between the axis of neighboring plain tubules
is not smaller than $\sim 1.3$ times the outer diameter of the cylinders.

The effective response of carbon nanotubes has been found to be sensitive to
both the inner cavity of hollow tubules and the anisotropy, which we have first
investigated with use of the MG approximation. Within this approach, we have
analyzed the mode strengths and positions of tangential and radial surface
plasmons, and have investigated the effect of the anisotropy and the
electromagnetic interactions between the tubes.

Finally, we have presented an analysis of the electric-field pattern
associated with the electromagnetic resonances that are present in the optical
absorption and the energy loss. We have shown that incident $p$-polarized
light impinging on a periodic array of hollow carbon nanotubes excites
tangential and radial plasmons at the surfaces of each tubule. In the case of
isolated hollow nanotubes, these surface plasmons are found to be prototypical
dipolar modes. In the close-packed regime, where nanotubes are nearly touching,
our calculations indicate that the incident radiation excites a very localized
tangential surface plasmon, while the radial plasmon is found to be very
similar to that observed in the case of isolated nanotubes.

\acknowledgements

J.M.P. gratefully acknowledges partial support by the University of the
Basque Country, the Basque Hezkuntza, Unibertsitate
eta Ikerketa Saila, and the Spanish Ministerio de Educaci\'on y Cultura.

\begin{figure}
\caption{Multishell nanotubes of inner and outer
radii $r$ and $R$, respectively, arranged in a square array with lattice
constant
$a$. The cylinders are infinitely long in the $y$ direction. The electromagnetic
interaction of
this structure with a normally incident plane wave of momentum ${\bf k}$
[$k_y=k_z=0$] and energy
$\omega$ is investigated.}
\end{figure}

\begin{figure}
\caption{Energy dependence of the real and imaginary parts of the principal
dielectric
functions, ${\rm Re}\,\varepsilon(\omega)$ and ${\rm
Im}\,\varepsilon(\omega)$, and the
energy-loss function, ${\rm Im}[-\varepsilon^{-1}(\omega)]$, for graphite,
taken
from Ref.\protect\onlinecite{Palik}\protect. Solid and dashed lines
represent
components perpendicular and parallel to the $c$-axis, respectively.}
\end{figure}

\begin{figure}
\caption{Imaginary part of the in-plane dipole polarizability per unit
volume of
isotropic plain ($\rho=0$) and hollow ($\rho\neq 0$) cylinders, as obtained
either
from Eq. (\ref{eq9}) with
$\varepsilon_\parallel(\omega)=\varepsilon_\perp(\omega)$ or from Eq.
(\ref{eq1p0}), as a function of frequency $\omega$ and for various values of the
ratio between
the inner and outer radii of the cylinders: $\rho=0, 0.2, 0.4, 0.6$, and
$0.8$.}
\end{figure}

\begin{figure}
\caption{Surface-mode positions: (a) $m_-,n_+$ and (b) $m_+,n_-$, as
obtained from
Eqs. (\ref{eq4pp}) and (\ref{eq5pp}) versus $\rho$, for various values of
the ratio
between the lattice constant and the outer diameter of the cylinders: $x=3$
(solid lines), $x=1.5$ (dotted lines),
and $x=1$ (dashed lines).}
\end{figure}

\begin{figure}
\caption{Strengths $B_\pm$, as obtained from Eq. (\ref{eq6pp}) versus
$\rho$, for
various values of the ratio between the lattice constant and the outer
diameter of the
cylinders: $x=3$ (solid lines), $x=1.5$ (dotted lines),
and $x=1$ (dashed lines).}
\end{figure} 

\begin{figure}
\caption{Imaginary part of the MG effective dielectric function and MG effective
energy-loss function of the periodic system described in Fig. 1, as
obtained from either Eqs. (\ref{eq11}) and (\ref{eq12}) with
$\varepsilon_\parallel(\omega)=\varepsilon_\perp(\omega)$ or from Eqs.
(\ref{eq1p}) and (\ref{eq2p}),
and for various values of the ratio between the inner and outer radii of
the cylinders:
$\rho=0$ (solid line), $\rho=0.2$ (dotted line), $\rho=0.4$ (short-dashed
line), $\rho=0.6$ (long-dashed line), and
$\rho=0.8$ (solid line). The ratio between the lattice constant and the
outer diameter of the cylinders is taken to be $x=1.5$.}
\end{figure}

\begin{figure}
\caption{Same as Fig. 3, as obtained from Eq. (\ref{eq9}) with use of
the actual
principal dielectric functions $\varepsilon_\perp(\omega)$ and
$\varepsilon_\parallel(\omega)$ of graphite.}
\end{figure}

\begin{figure}
\caption{Same as Fig. 6, as obtained from Eqs. (\ref{eq11}) and
(\ref{eq12}) with
use of the actual principal dielectric functions $\varepsilon_\perp(\omega)$ and
$\varepsilon_\parallel(\omega)$ of graphite.}
\end{figure}

\begin{figure}
\caption{The real and imaginary parts of the long-wavelength effective
dielectric
function, ${\rm Re}\,\varepsilon_{eff}(\omega)$ and ${\rm
Im}\,\varepsilon_{eff}(\omega)$, and the energy-loss function, ${\rm
Im}[-\varepsilon_{eff}^{-1}(\omega)]$, of a periodic array of plain
($\rho=0$) carbon
nanotubes, for $p$ polarized electromagnetic excitations. Solid, long-dashed,
and short-dashed lines represent our full calculated results, as obtained from
Eq. (\ref{eq3}) for ratios between the lattice constant and the outer diameter of
the cylinders
$x=2$, $x=1.5$, and $x=1.3$, respectively. The dotted lines represent MG
results, as
obtained from Eqs. (\ref{eq11}) and (\ref{eq12}).}
\end{figure}

\begin{figure}
\caption{Same as Fig. 9, for ratios between the lattice constant and the
outer diameter of
the cylinders $x=1.1$ (solid line) and $x=1.03$ (dashed line).}
\end{figure}

\begin{figure}
\caption{The intensity of the electric field and the 
corresponding charge density
generated
by a normally incident $p$ polarized electromagnetic plane wave impinging on a
periodic system of
hollow ($\rho=0.6$) carbon nanotubes with $x=2.0$. Both the electric field and
the induced charge have been evaluated at two different frequencies: 
$\omega_1=5.2\,{\rm eV}$ (the frequency
at which the optical absorption is maximum) and $\omega_2=5.9\,{\rm eV}$ 
(the frequency at which the energy loss is maximum). White/dark areas mean small/large values of the electric-field intensity and the induced charge.}
\end{figure}

\begin{figure}
\caption{Same as Fig. 11, for $x=1.03$, $\omega_1=4.75\,{\rm eV}$, 
and $\omega_2=6.1\,{\rm eV}$.} 
\end{figure}

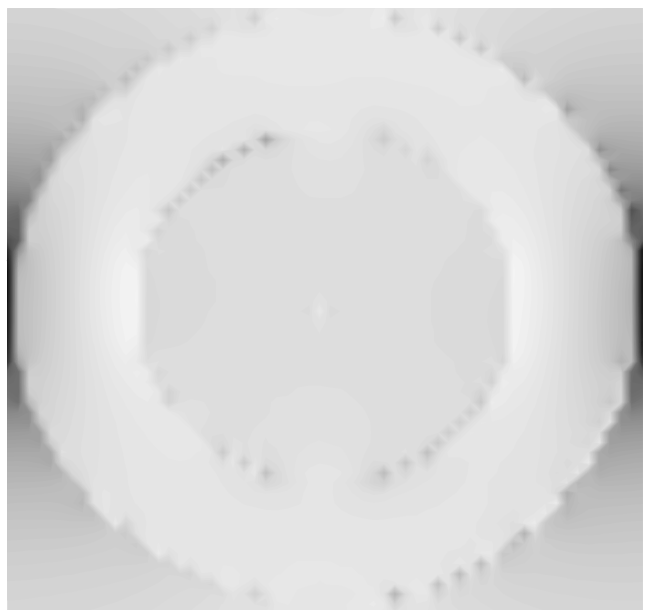
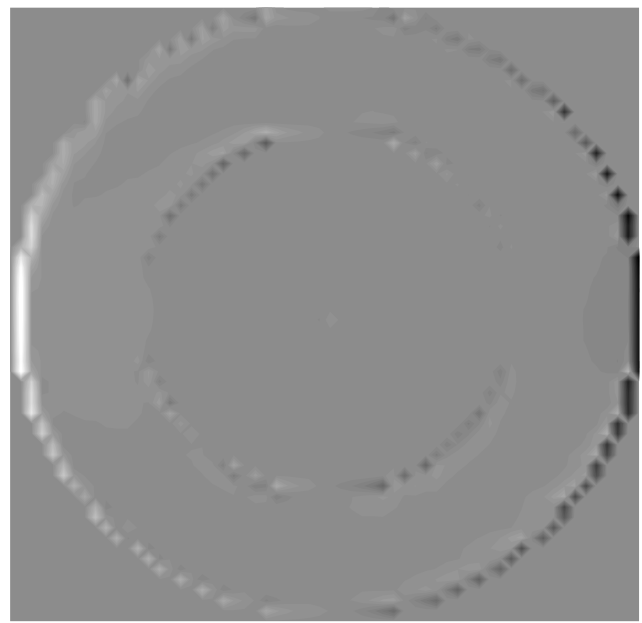
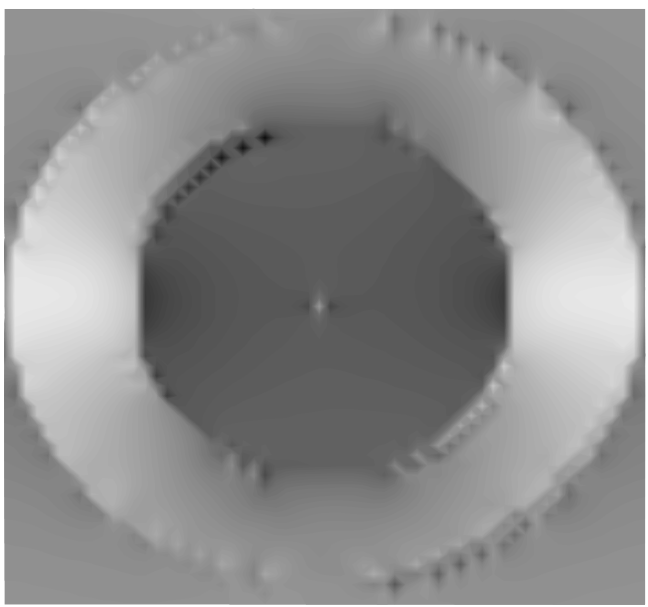
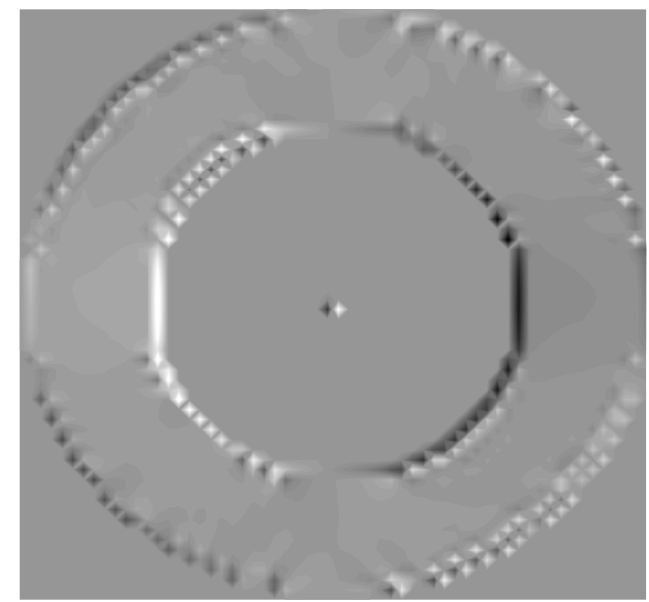
\end{document}